# Quantum Anomalous Hall Effect in Magnetic Doped Topological Insulators and Ferromagnetic Spin-Gapless Semiconductors – A Perspective Review

*Muhammad Nadeem*[*1,2], *Alex R. Hamilton*[3,4], *Michael S. Fuhrer*[5,6], *Xiaolin Wang*[*1,2]
[1]Institute for Superconducting and Electronic Materials (ISEM), Australian Institute for Innovative Materials (AIIM), University of Wollongong, Wollongong, New South Wales 2525, Australia
[2]ARC Centre of Excellence in Future Low-Energy Electronics Technologies (FLEET), University of Wollongong, Wollongong, New South Wales 2525, Australia
[3]School of Physics, University of New South Wales, Sydney, NSW 2052, Australia
[4]ARC Centre of Excellence in Future Low-Energy Electronics Technologies, University of New South Wales, Sydney, NSW 2052, Australia.
[5]School of Physics and Astronomy, Monash University, Monash, Victoria 3800, Australia
[6]ARC Centre of Excellence in Future Low-Energy Electronics Technologies (FLEET), Monash University, Monash, Victoria 3800, Australia
E-mail: (( muhammad.nadeem@seecs.edu.pk and xiaolin@uow.edu.au))

Quantum anomalous Hall effect, with a trademark of dissipationless chiral edge states for electronics/spintronics transport applications, can be realized in materials with large spin-orbit coupling and strong intrinsic magnetization. After Haldane's seminal proposal, several models have been presented to control/enhance the spin-orbit coupling and intrinsic magnetic exchange interaction. After brief introduction of Haldane model for spineless fermions, following three fundamental quantum anomalous Hall models are discussed in this perspective review: *(i)* low-energy effective four band model for magnetic-doped topological insulator $(Bi,Sb)_2Te_3$ thin films, *(ii)* four band tight-binding model for graphene with magnetic adatoms, and *(iii)* two (three) band spinfull tight-binding model for ferromagnetic spin-gapless semiconductors with honeycomb (kagome) lattice where ground state is intrinsically ferromagnetic. These models cover two-dimensional Dirac materials hosting spinless, spinful and spin-degenerate Dirac points where various mass terms open a band gap and lead to quantum anomalous Hall effect. With emphasize on the topological phase transition driven by ferromagnetic exchange interaction and its interplay with spin-orbit-coupling, we discuss various symmetry constraints on the nature of mass term and the materialization of these models. We hope this study will shed light on the fundamental theoretical perspectives of quantum anomalous Hall materials.

## 1. Introduction

Qauntum anamolous Hall (QAH) effect[1] referes to the quantization of anamolous Hall (AH) conductivity due to coupling between intrinsic magnetization and spin-orbit coupling (SOC). Unlike spin-degenerate chiral edge states in integer quantum Hall effect where time-reversal symmetry (TRS) is broken through external magnetic field, the trademark of QAH effect is the spinless/spinful chiral edge states. Such macroscopic quantum phenomenon, originating from the bulk topology of QAH systems, is ideal for transport in both electronics and spintronics devices.

After the sminal proposal by Haldane[1], a number of theoretical models were proposed for the realization of QAH effect in realistic condensed mater systems sharing the same key feature – achieving simultanous controle over intrinsic magnetization and SOC. This aim is associated with two decisive energy scales for the observation of QAH effect: association of the SOC with the bulk topological band gap and the association of ferromagnetism with the Curie temperature. The smaller of these two energy scales defines the upper limit for the observable temperature of QAH effect. As a result, to assess the realistic temperature scale for QAH effect, the necessity of a delicate interplay between these two energy scales plays the key role in the search of suitable materials.

Keeping this key feature for the materialization of QAH effect in view, several proposals for QAH effect have been made where intrinsic magnetization and SOC can be tunned through the proximity effect or the induction of magnetic impurities. For example, QAH effect by inducing magnetic impurities in topological insulator $(Bi,Sb)_2Te_3$ thin films[2-14], QAH effect in hybrid class of materials where monolayers of Van der Waals materials are doped with transition metal (TM) adatoms such as graphene with 5d-adatoms[15] and $MoS_2$ monolayer with 3d-adatoms[16]. Similar to magnetic adatoms on non-magnetic Dirac materials[15, 16], enhanced SOC for QAH effect can be realized by depositing atomic layers of heavy elements on the surface of a magnetic insulator (MnX; X=S, Se, Te)[17]. Moreover, successful fabrication of Van der Waals heterostructures[18-20] motivates the possibility of QAH effect in various heterostructures such as TM-oxide based heterostructures[21-25], ferromagnetic insulator based heterostructures with both topological insulator[26-28] and magnetic-doped topological insulators[29-31], graphene-based heterostructures with both ferromagnetic[32-36] and antiferromagnetic materials[37, 38]. However, to date, the QAH effect has only been materialized in ferromagnetic topological insulators, $Cr^{9-12}$ and $V^{13}$ doped $(Bi,Sb)_2Te_3$ thin films.

Despite these proximity/impurity induced systems for QAH effect, there are even simpler condensed matter systems – ferromagnetic spin-gapless semiconductors (SGS)[39, 40] with a signature of spinful Dirac dispersion in the absence of SOC. For example, QAH effect in monolayers of TM-compounds such as TM-halides[41-47], TM-oxides[48-50], TM-organic framework[51-53] and TM-compounds with $d$-electron kagome lattices[54-58]. The recent discovery of intrinsic ferromagnetism in two-dimensionally thin TM-compounds[59-62], opens a new avenue for the experimental realization of QAH effect in ferromagnetic SGS where both ferromagnetism and SOC are intrinsically present.

There are a number of excellent reviews covering from fundamental concepts, physical origin for non-zero Chern number, and the list of theoretically predicted and experimentally realized condensed matter systems for QAH effect[62-73]. Our aim is not the search for exhaustive list of materials but rather the discussion of key features and the interplay required between ferromagnetic exchnage interaction and SOC in materials that can host QAH effect. In this perspective review, after brief introduction of Haldane model for spinless fermions, following three types of quantum anomalous Hall models are discussed: *(i)* four-band low-energy effective model for magnetic-doped topological insulator $(Bi,Sb)_2Te_3$ thin films[2], *(ii)* four band tight-binding model for graphene with magnetic adatoms[74-76], and *(iii)* two (three) band spinfull tight-binding model for ferromagnetic SGS with honeycomb[51] and kagome[54, 55] lattices where ground state is intrinsically ferromagnetic. These spinful models can be thought of as single copy of four (six) band tight-binding model for quantum spin Hall insulators[77-80] in monolayers with honeycomb (kagome) lattice.

These models cover two-dimensional Dirac materials hosting spinless, spinful, spin-degenerate Dirac points where various mass terms open band gap and lead to QAH effect. Subject to various symmetry constraints, the nature of mass term is quite different in these models. With emphasize on the topological phase transition driven by ferromagnetic exchange interaction and its interplay with SOC, it can be envisaged that the materialization of these models is subject to various constraints. In type (*i*) models, where it is known that SOC is inherently strong in $(Bi,Sb)_2Te_3$ thin films, ferromagnetic exchange interaction is induced through magnetic ion doping or proximity effect and hence a more delicate interplay among SOC and exchange interaction is required such that magnetization do not affect the band topology. In type (*ii*) models, topologically nontrivial band gap is subject to combined effect of exchange interaction induced by TM adatoms and enhanced SOC through heavy elements adatoms. In the type (*iii*) models, since the ground state is intrinsically ferromagnetic, the only relevant task is the search of SGS with large intrinsic SOC.

In principle, materialization is possible in all three types of models, however, quantum anomalous Hall effect has only been experimentally demonstrated so far in (Bi,Sb)$_2$Te$_3$ thin films. Proximity/impurity induced ferromagnetism in (Bi,Sb)$_2$Te$_3$ thin films shows very rich behaviour and leads to QAH effect whether the system is originally in topologically trivial or non-trivial phase. On the other hand, proximity/impurity induced ferromagnetism drives Kane-Mele type spin-orbit coupled QSH insulators directly into a metallic phase. Despite this topological phase transition without entering QAH effect, Kane-Mele type next-nearest neighbour tunnelling can play an important role in the materialization of QAH effect in ferromagnetic SGS. We hope this study will shed light on the fundamental theoretical perspectives of QAH materials.

## 2. Haldane Model

In 1988, Haldane proposed a seminal tight binding model for graphene where next-nearest neighbour (NNN) hopping with complex matrix elements breaks time reversal symmetry (TRS) and induces a mass term which leads to QAH effect without Landau levels[1]

$$H = -t \sum_{\langle ij \rangle} c_i^\dagger c_j - m_H \sum_{\langle\langle ij \rangle\rangle} e^{i v_{ij} \varphi} c_i^\dagger c_j + m_S \sum_{i\lambda} c_i^\dagger c_i \qquad (1)$$

Here $c_{i\alpha}^\dagger (c_{i\alpha})$ is the creation (annihilation) electron operator on site $i$, $t$ is the nearest-neighbour (NN) hoping, $m_H$ is the Haldane parameter where $\varphi$ is the phase attained by electron via NNN hopping, and $m_S = +1(-1)$ for A(B) sublattices is the Semenoff[81] parameter which induces staggered sublattice potential. If $\varphi = 0$, this model reduces to the Wallace model[82] where the second term becomes NNN hopping with real matrix elements and does not open a band gap but simply shifts the Fermi level and leads to a trivial metal. When $m_H = 0$, the model represents Semenoff type[81] trivial insulating system with a mass gap of $2m_S$.

For infinite sheet of honeycomb lattice, where $\mathbf{d}_i$ and $\mathbf{v}_i$ represent the NN and NNN bond vectors, the Haldane model transforms to a two-band Bloch Hamiltonian in momentum space with basis $\psi_k \equiv (a_k, b_k)$

$$H(\mathbf{k}) = \begin{pmatrix} h_+(\mathbf{k}) + m_s & h_t(\mathbf{k}) \\ h_t^*(\mathbf{k}) & h_-(\mathbf{k}) - m_s \end{pmatrix} \qquad (2)$$

Here $h_\pm(k) = -2m_H \sum_{i=1}^{3} \cos(\mathbf{k} \cdot \mathbf{v}_i \pm \varphi)$ and $h_t(k) = -t \sum_{i=1}^{3} e^{-i\mathbf{k} \cdot \mathbf{d}_i}$. In the long wavelength limit, the low energy single-particle band dispersion in the vicinity of Dirac points reads

$$E(\mathbf{q}) = 3m_H \cos(\varphi) + \eta_{v/c} \sqrt{v_F^2 |\mathbf{q}|^2 + \left| m_S - 3\sqrt{3}\eta_\tau m_H \sin(\varphi) \right|^2} \qquad (3)$$

where $v_F = 3at/2$ is Fermi velocity and the index $\eta_v = -\eta_c = -1$ represents valence/conduction bands. In the absence of both Haldane and Semenoff terms, the NN term shows spinless Dirac dispersion. When inversion symmetry is present, the Haldane term opens a symmetric topologically nontrivial Dirac gap of $E_g = 6\sqrt{3} m_H \sin(\varphi)$ which leads to chiral edge states. However, with broken inversion symmetry, band gap is different at different valleys: while the band gap remain positive and increases with increasing $m_S$ at valley K', the band gap decreases at valley K, closes when $m_S = 3\sqrt{3} m_H \sin(\varphi)$, and the system enters into trivial regime when $m_S > 3\sqrt{3} m_H \sin(\varphi)$. The bulk band evolution and the corresponding energy dispersion for semi-infinite nanoribbon showing chiral edge states is depicted in **Figure 1**.

For a general two-band model described by Bloch Hamiltonian $H(\mathbf{k}) = \mathbf{h}(\mathbf{k}) \cdot \boldsymbol{\sigma}$, the Chern number can be written in terms of Hamiltonian unit vector $\hat{\mathbf{h}}(\mathbf{k})$.

$$C = \frac{1}{4\pi} \int d\mathbf{k} \left[ \hat{\mathbf{h}} \cdot (\partial_i \hat{\mathbf{h}} \times \partial_j \hat{\mathbf{h}}) \right] \qquad (4)$$

That is, the Chern number is equivalent to the number of times $\hat{\mathbf{h}}(\mathbf{k})$ wraps around the unit sphere in momentum space. For the Haldane model (1), it is straightforward to show that

$$C = \frac{1}{2}\left[ \text{sgn}(m_+) - \text{sgn}(m_-) \right] \qquad (5)$$

where $m_\pm$ are the mass terms at valleys $K_\pm$. Since the Haldane mass term carries different signs at different valleys and each valley contributes C = 1/2, the total Chern number remains C = 1 and leads to quantized Hall conductance. Berry curvature distribution and phase diagram for Haldane model is shown in Figure 1.

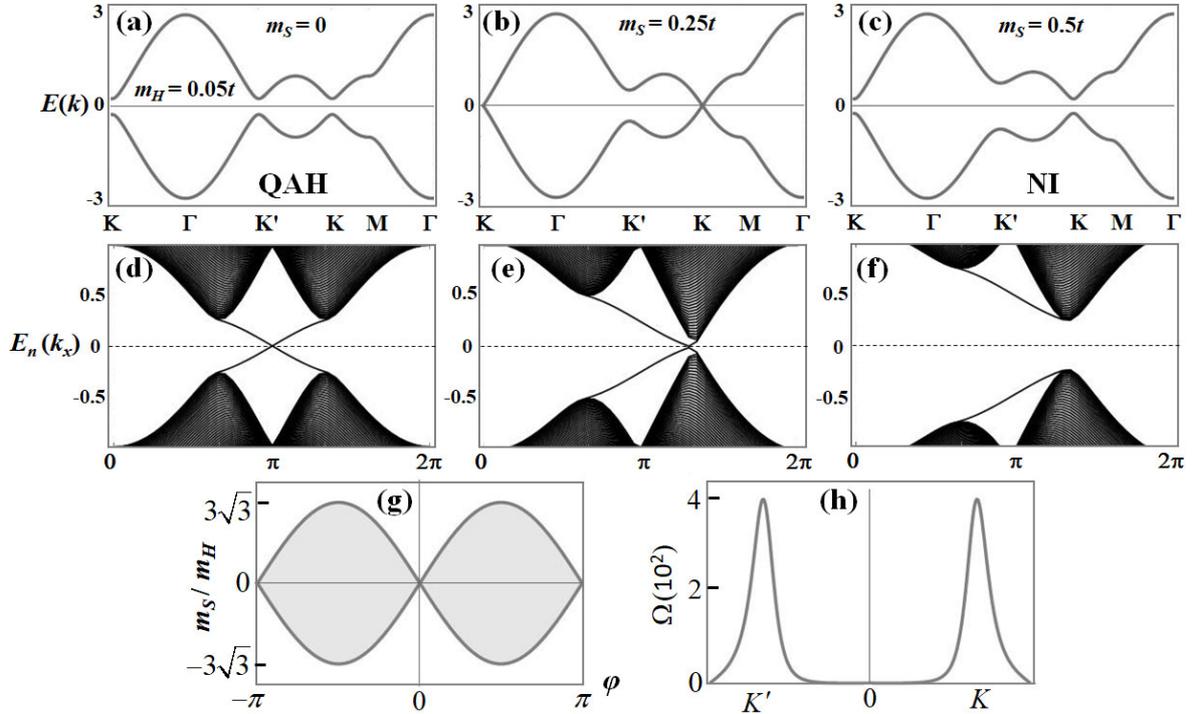

Figure 1: Haldane model for QAH effect: **(a-c)** QAH effect induced by next-nearest neighbour tunnelling **(a)**, gapless phase violating fermion doubling theorem **(b)** and trivial insulator **(c)**. **(d-f)** Band dispersion for semi-infinite honeycomb sheet with zigzag edges. **(g, h)** Phase diagram and Berry curvature distribution for Haldane model. Here the filled area of the phase diagram represents the topologically nontrivial regime with Chern number C = ±1.

### 3. QAH Effect in Magnetic Doped (Bi,Sb)$_2$Te$_3$ Thin Films

Inherently strong SOC in topological insulators motivates the realization of QAH phase in magnetic topological insulators by inducing ferromagnetism through impurity induction or proximity effect. In this section, we discuss the low energy effective model for topological insulator (Bi,Sb)$_2$Te$_3$ thin films[83, 84] where TRS is broken though Zeeman-type exchange field[2]. This model captures the true essence of interplay between magnetization and SOC for QAH effect though topological phase transition driven by Zeeman exchange interaction. The low energy effective four-band model Hamiltonian for topological insulator $(Bi,Sb)_2Te_3$ thin films[83, 84] can be written in the basis $|+\uparrow\rangle, |-\downarrow\rangle, |+\downarrow\rangle, |-\uparrow\rangle$ as

$$H(\mathbf{q}) = \begin{pmatrix} m(\mathbf{q}) & iv_F q_- & 0 & 0 \\ -iv_F q_+ & -m(\mathbf{q}) & 0 & 0 \\ 0 & 0 & m(\mathbf{q}) & -iv_F q_+ \\ 0 & 0 & iv_F q_- & -m(\mathbf{q}) \end{pmatrix} \quad (6)$$

where $|\pm\uparrow\rangle = (|t\uparrow\rangle \pm |b\uparrow\rangle)/\sqrt{2}$, $|\pm\downarrow\rangle = (|t\downarrow\rangle \pm |b\downarrow\rangle)/\sqrt{2}$, $t(b)$ represent the top(bottom) surface states, and $\uparrow(\downarrow)$ represent the spin-up (spin-down) surface modes. Here $q_\pm = q_x \pm q_y$, $v_F = A/\hbar$ is the surface Fermi velocity, and $m(\mathbf{q}) = m_0 - m_1(q_x^2 + q_y^2)$ is the tunnelling between top and bottom surfaces. Here $A$, $m_0$, and $m_1$ are the material parameters. As shown in Figure 2(a) and 2(d), the four band model $H(\mathbf{q})$ represents QSH phase with inverted band dispersion for $m_0 m_1 > 0$, gapless state for $m_0 = 0$, and trivial insulator for $m_0 m_1 < 0$. When Zeeman exchange interaction is introduced, the model reads[2]

$$H(\mathbf{q}, m_z) = \begin{pmatrix} m(\mathbf{q}) + m_z & iv_F q_- & 0 & 0 \\ -iv_F q_+ & -m(\mathbf{q}) - m_z & 0 & 0 \\ 0 & 0 & m(\mathbf{q}) - m_z & -iv_F q_+ \\ 0 & 0 & iv_F q_- & -m(\mathbf{q}) + m_z \end{pmatrix} \quad (7)$$

The bulk band dispersion, near $\Gamma$-point, for upper $H_+(\mathbf{q}, m_z)$ and lower block $H_-(\mathbf{q}, m_z)$ are

$$E_\pm(\mathbf{q}) = \eta_{v/c} \sqrt{v_F^2 |\mathbf{q}|^2 + |m(\mathbf{q}) \pm m_z|^2} \quad (8)$$

In the presence of ferromagnetic exchange interaction, the system $H(\mathbf{q}, m_z)$ transforms to QAH phase irrespective of whether *a prior* state of $H(\mathbf{q})$ is normal insulator or QSH insulator as shown in Figure 2. In both of these regimes, the basic mechanism that leads to QAH phase is the appearance (disappearance) of exchange interaction induced (suppressed) band inversion in only one of the two time reversal partners with opposite chirality: $H_+(\mathbf{q})$ and $H_-(\mathbf{q})$.

Let's assume $m_1 > 0$. A similar situation holds for $m_1 < 0$ while tunning $m_z$ in opposite fashion to the one we discuss here. *Case-I:* Suppose $H(\mathbf{q})$ is originally in the topologically trivial phase, $m_0 > 0$. When Zeeman exchange interaction is introduced, $H(\mathbf{q}, m_z)$ is a trivial insulator when $|m_0| > |m_z|$ and QAH insulator with $C = \pm 1$ when $|m_0| < |m_z|$. *Case-II:* Suppose $H(\mathbf{q})$ is originally in the inverted regime, $m_0 m_1 < 0$. When Zeeman exchange interaction is introduced, $H(\mathbf{q}, m_z)$ remains a QSH insulator for $|m_0| > |m_z|$ and transforms to QAH insulator with $C = \pm 1$ when $|m_0| < |m_z|$. The phase diagram and the bulk band structure for both of these cases and the topological phase transition driven by magnetization $m_z$ are shown in Figure 2.

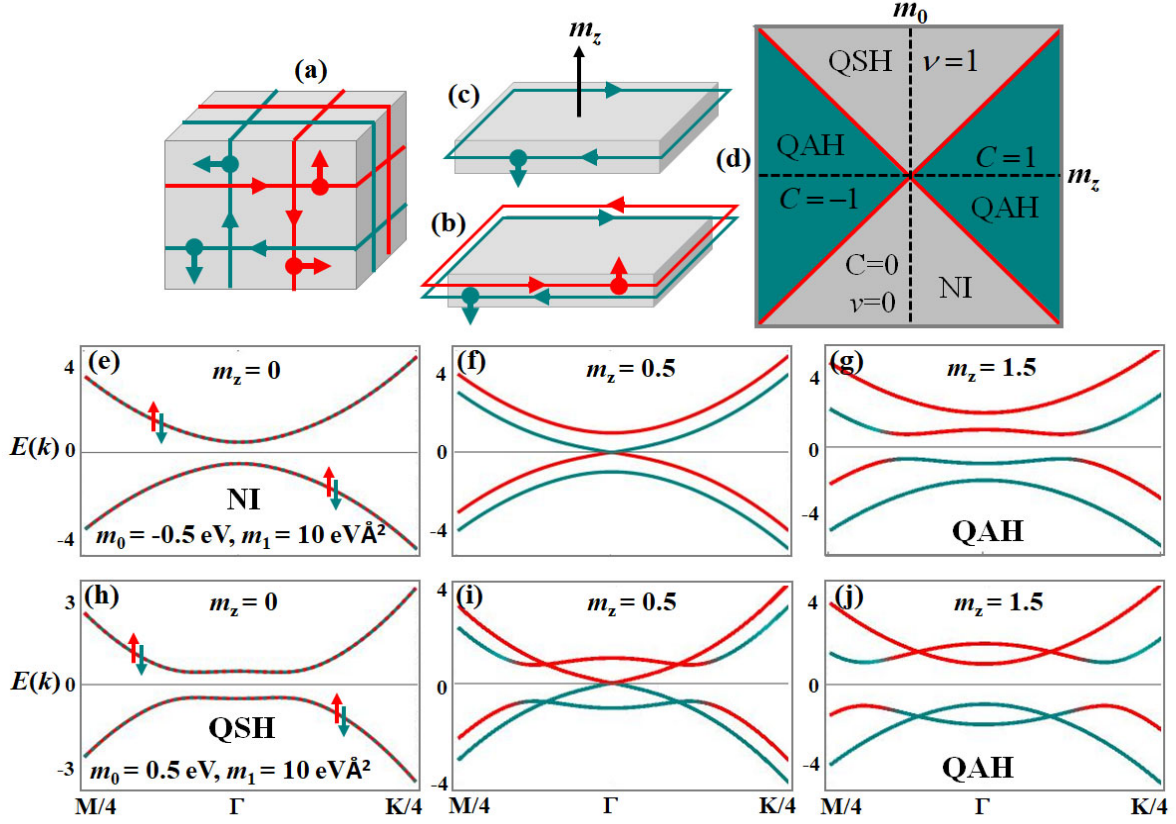

**Figure 2: Topological phase transition in ferromagnetic thin films of topological insulators. (a-d)** Three-dimensional topological insulator with single Dirac cone on each surface **(a)**, thin films with helical **(b)** and chiral **(c)** edge states, and phase diagram **(d)** representing normal insulator (NI), QSH insulator, and QAH insulator. The red lines, represent the critical gapless phase. **(e-g)** Topological phase transition driven by Zeeman exchange interaction, when the system is in trivial phase **(e)**. **(h-j)** Topological phase transition driven by Zeeman exchange interaction, when the system is in QSH phase **(h)**. In both of these cases, Zeeman exchange interaction leads to critical point at $|m_z| = |m_0|$ **(f, i)** and then QAH phase when $|m_z| > |m_0|$ **(g, j)**. We set $A = 2.5$ eVÅ here.

## 4. QAH Effect in Graphene via Exchange and Rashba Effect

Apart from the Haldane model, where the leading term is the NNN tunnelling, QAH effect can also be realized in graphene via the combined effects of Rashba SOC modelled via NN tunnelling and ferromagnetic exchange interaction induced by adatoms[74-76]. Tight binding model Hamiltonian for graphene with Rashba SOI and out-of-plane Zeeman magnetization reads

$$H = t \sum_{\langle ij \rangle \alpha} c_{i\alpha}^\dagger c_{j\alpha} + i\lambda_{ER} \sum_{\langle ij \rangle \alpha\beta} c_{i\alpha}^\dagger (\mathbf{s}_{\alpha\beta} \times \hat{\mathbf{d}}_{ij})_z c_{j\beta} + m_z \sum_{i\alpha} c_{i\alpha}^\dagger s^z c_{i\alpha} \qquad (9)$$

Here, first term is the NN hopping, second term is the Rashba SOC associated with broken out-of-plane mirror symmetry, and the third term is ferromagnetic exchange interaction. In the long wavelength limit, diagonalization of low energy effective case gives the energy band dispersion which remains same at both valleys and reads

$$E(\mathbf{q}) = \eta_{v/c} \sqrt{v_F^2 |\mathbf{q}|^2 + 2\Delta_{ER}^2 + m_z^2 + 2\eta_s \sqrt{v_F^2 |\mathbf{q}|^2 \Delta_{ER}^2 + v_F^2 |\mathbf{q}|^2 m_z^2 + \Delta_{ER}^4}} \qquad (10)$$

In the presence of Rashba SOC, spin is not a good quantum number and $\eta_s = +1(-1)$ represent spin chirality of angular momentum eigenstates. It is known that neither exchange interaction nor Rashba SOC can open band gap solely. However, when both Rashba SOI and exchange interaction are simultaneously present, a nontrivial band gap opens. As shown in Figure 3, exchange interaction leads system into inverted regime such that spin-degenerate circles at energy $\varepsilon = 0$ are formed in momentum space. Along these spin-degenerate circles, Rashba SOC mixes up and down-spins, induces an avoided band crossing, and opens a band gap around these circles. As a result, the combined effect of Rashba SOC and exchange interaction leads to avoided crossings at momentum $q_\Delta$ where the absolute band gap reads

$$E_\Delta(q_\Delta) = \frac{2m_z \Delta_{ER}}{\sqrt{m_z^2 + \Delta_{ER}^2}}, \quad q_\Delta = \frac{m_z \sqrt{m_z^2 + 2\Delta_{ER}^2}}{v_F \sqrt{m_z^2 + \Delta_{ER}^2}} \tag{11}$$

It has been explicitly shown[74-76] that the induced band gap is topologically nontrivial and leads to Chern number $C = 2$. Unlike Haldane model and QAH in spinful ferromagnetic semiconductors (to be discussed section 6), the leading SOC term in this model is Rashba SOC. However, the QAH phase remains intact even in the presence of intrinsic atomic SOC as long as it smaller than the Rashba SOC.

Since spin is not a good quantum number due to presence of Rashba SOC, the nontrivial band topology can be verified by direct calculation of Chern invariant and skyrmion-like spin texture via knowledge of Bloch wavefunctions of filled bands. That is, Chern invariant can be written in terms of Brillouin-zone integrals over momentum-space Berry curvature as[85, 86]

$$C = \frac{1}{2\pi} \sum_n \int d^2k (\Omega_n)_z \tag{12}$$

Here $\mathbf{A}_n(\mathbf{k}) = i\langle u_n(\mathbf{k})|\nabla_\mathbf{k}|u_n(\mathbf{k})\rangle$ and $\mathbf{\Omega}_n(\mathbf{k}) = (\nabla_\mathbf{k} \times \mathbf{A}_n(\mathbf{k})) = i\langle \nabla_\mathbf{k} u_n(\mathbf{k})|\times|\nabla_\mathbf{k} u_n(\mathbf{k})\rangle$ are Berry connection and Berry curvature for the Bloch wavefunction of $n$th band $|u_n(\mathbf{k})\rangle$ and the summation runs over $n$, the band index representing number of bands below Fermi level. As shown in Figure 2(f), non-vanishing Berry curvature distribution in the vicinity of each Dirac point leads to Chern invariant $C = 1$ and hence total Chern invariant of $C = 2$ when Berry curvature is integrated over whole Brillouin zone.

In addition, direct evaluation of spin components for filled Bloch bands shows that the most of contribution of non-vanishing Berry curvature comes for low energy filled bands. As shown in Figure 3, the spin texture of low energy filled band at both valley K and K' is skyrmion-like [87]: spin is pointing towards south Pole at the valleys (q=0) while towards north pole away from valleys. Moreover, the spin flip circular regime can be clearly seen at the equator where total spin is in-plane. On the other hand, the spin texture of high energy filled bands is topologically trivial at both valleys - spins are uniformly pointing toward the South Pole in the whole momentum space, and hence contribution form high energy filled bands in topological charge is negligibly small. As a result, the contribution to the total Chern number of $C = 2$ – where each skyrmion at valley K and K' contributes to a Chern number of $C=1$[88] – is mostly from the low energy filled band.

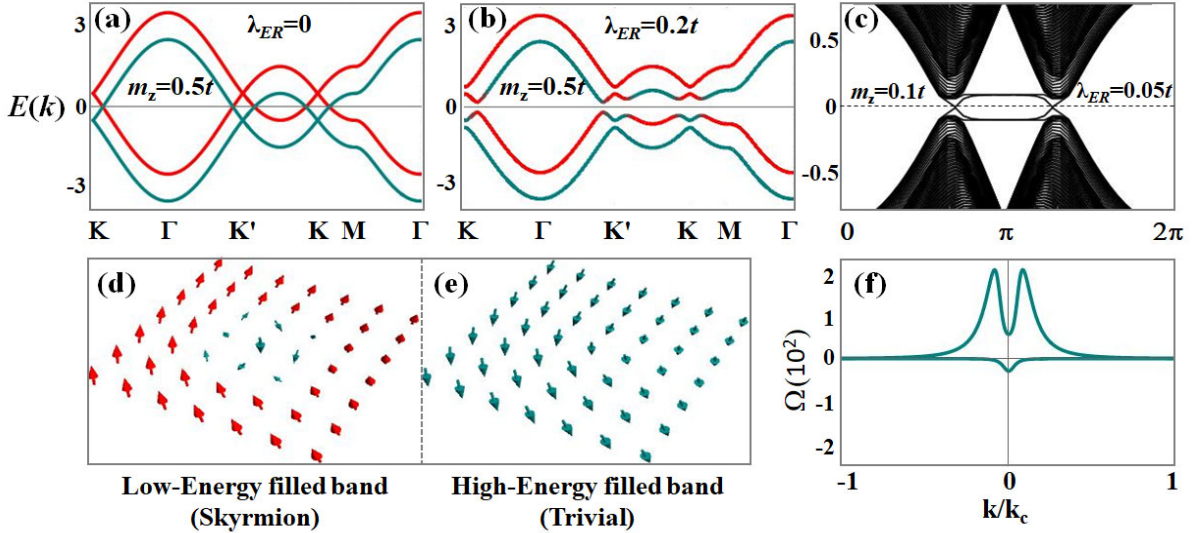

**Figure 3: QAHE induced by Rashba SOI and Zeeman exchange interaction in graphene. (a-c) Band structure**. Bulk band splitting due to Zeeman exchange interaction **(a)** and the effect of Rashba SOI **(b)**. One-dimensional band dispersion for semi-infinite sheet of honeycomb nanoribbon with zigzag edges **(c)** confirms that the bulk band gap is topologically nontrivial and is characterized by Chern number $C = 2$. **(d-f) Spin texture, and Berry curvature distribution at valley K.** Spin texture for low energy filled band **(d)** shows topologically nontrivial skyrmion-like behaviour while that for high energy filled band **(e)** is trivial. Non-vanishing Berry curvature distribution in the vicinity of Dirac point leads to non-zero Chern invariant. Both the spin texture and Berry curvature distribution show similar behaviour at valley K'.

### 5. Magnetic Doped QSH Insulators in Honeycomb and Kagome Monolayers

Similar to magnetic doped topological insulator $(Bi,Sb)_2Te_3$ thin films[2], we discuss the tight binding model for QSH insulators in honeycomb[77, 78] and kagome lattice[79, 80] with Kane-Mele type SOC[77, 78] and magnetic exchange interaction. When Zeeman-type exchange field is added, a general tight binding model for QSH insulator in honeycomb/kagome lattices with Kane-Mele type SOC takes the form

$$H = -t\sum_{\langle ij\rangle\alpha} c_{i\alpha}^\dagger c_{j\alpha} + i\lambda_0 \sum_{\langle\langle ij\rangle\rangle\alpha\beta} v_{ij} c_{i\alpha}^\dagger s_{\alpha\beta}^z c_{j\beta} + \sum_{i\alpha} c_{i\alpha}^\dagger (\varepsilon_0 + m_z s^z) c_{i\alpha} \quad (13)$$

Here $c_{i\alpha}^\dagger (c_{i\alpha})$ is the creation (annihilation) electron operator with spin polarization $\alpha = \uparrow,\downarrow$ on site $i$, Pauli matrix $s^z$ describes electron intrinsic spin while $s_{\alpha\beta}^z$ are the corresponding matrix elements describing spin polarization $\alpha$ and $\beta$ on sites $i$ and $j$, and $v_{ij} = \mathbf{d}_{ik} \times \mathbf{d}_{kj} = \pm 1$ connects sites $i$ and $j$ on sublattice A (B) via the unique intermediate site $k$ on sublattice B (A). Here $\mathbf{d}_{ik}$ and $\mathbf{d}_{kj}$ are the nearest-neighbour bond vectors connecting A and B sublattices. First term is the nearest neighbour hoping with amplitude $t$, second term is the Kane-Mele type SOC with strength $\lambda_0$, while third term incorporates both on-site energy $\varepsilon_0$ and impurity/proximity induced Zeeman exchange interaction with strength $m_z > 0$. For simplicity, we ignore the spin mixing Rashba type SOC term here.

**5.1. Four-band model for honeycomb lattice:** For infinite sheet of honeycomb lattice, where $\mathbf{d}_i$ and $\mathbf{v}_i$ represent the nearest neighbour and next-nearest neighbour bond vectors, model (12) transforms to four-band Bloch Hamiltonian in momentum space with basis $\psi_k \equiv (a_{k,\uparrow}, b_{k,\uparrow}, a_{k,\downarrow}, b_{k,\downarrow})$

$$H(\mathbf{k}) = \begin{pmatrix} h_\lambda(k) + m_z & h_t(k) & 0 & 0 \\ h_t^*(k) & -h_\lambda(k) + m_z & 0 & 0 \\ 0 & 0 & -h_\lambda(k) - m_z & h_t(k) \\ 0 & 0 & h_t^*(k) & h_\lambda(k) - m_z \end{pmatrix} \quad (14)$$

Here $h_\lambda(k) = -2\lambda_0 \sum_{i=1}^{3} \sin(\mathbf{k} \cdot \mathbf{v}_i)$ and $h_t(k) = -t \sum_{i=1}^{3} e^{-i\mathbf{k} \cdot \mathbf{d}_i}$. In the long wavelength limit, low energy single-particle band dispersion in the vicinity of Dirac points reads

$$E(\mathbf{q}) = \eta_s m_z + \eta_{v/s} \sqrt{v_F^2 |\mathbf{q}|^2 + |3\sqrt{3}\lambda_0|^2} \quad (15)$$

where $\eta_s = +(-)$ for spin up(down) sector and $v_F = 3at/2$ is Fermi velocity. In the presence of inversion and time reversal symmetries, SOC opens a symmetric topological Dirac gap of $E_g = 6\sqrt{3}\lambda_0$. When Zeeman exchange interaction is activated, it leads to a topological phase transition from QSH phase for $0 < m_z < 3\sqrt{3}\lambda_0$ to regular ferromagnetic metal when $m_z > 3\sqrt{3}\lambda_0$. The evolution of bulk band structure with increasing magnetization strength shows that the QSH insulator can be viewed as two copies of QAH phase separated along the energy axis. However, the system remains in metallic phases. If the Fermi level lies within Dirac gap of one of the spin sectors while the other spin sector is removed, it gives QAH effect with chiral edge state. Bulk band dispersion and corresponding energy dispersion for semi-infinite nanoribbon are shown in Figure 4.

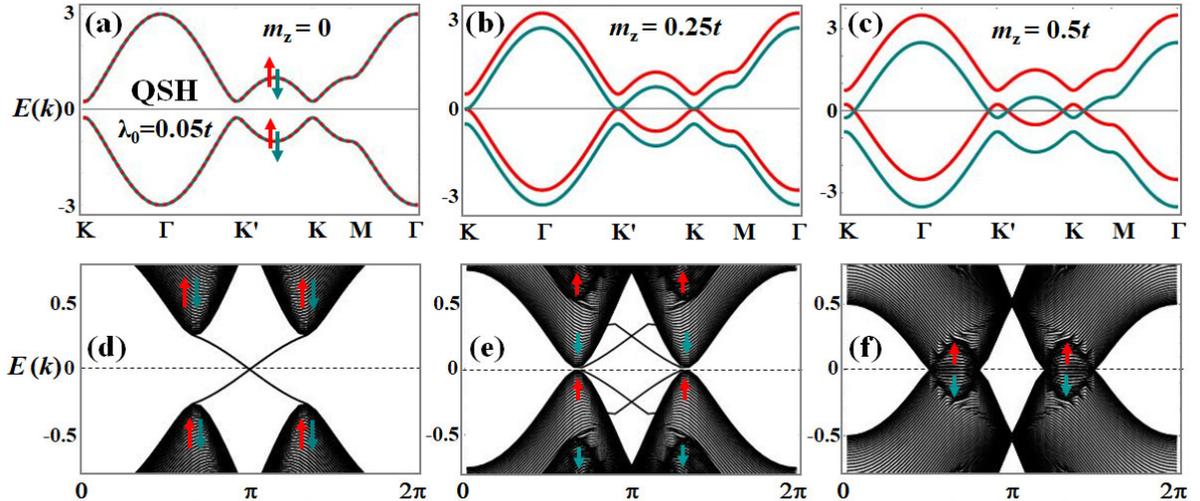

**Figure 4: Topological phase transition, driven by Zeeman exchange interaction, in QSH system with honeycomb lattice.** (*Top panel*) Bulk band structure for infinite sheet of honeycomb lattice, showing topological phase transition driven by Zeeman exchange interaction where QSH phase **(a)** can be seen as two QAH-like spinful copies separated along energy axis but the Zeeman exchange interaction turns the system into a regular ferromagnetic metal **(b, c)**. (*Bottom panel*) Corresponding energy dispersion for semi-infinite sheet of honeycomb lattice with zigzag edges showing helical edge states in QSH phase **(d)** and regular ferromagnetic metal **(e, f)**.

**5.2. Six-band model for kagome lattice:** Similar to the honeycomb lattice, the kagome lattice hosts Dirac dispersion near the corners of Brillion zone. However, unlike graphene, there are extra flat bands which are degenerate with dispersing bands at the centre of the Brillion zone. When Kane-Mele type SOC is activated, it opens two topological gaps; one between dispersing bands at the valleys K and K' and the other between flat bands and dispersing bands at Γ-point. For infinite sheet of kagome lattice, where $\mathbf{d}_i$ and $\mathbf{v}_i$ represent the NN and NNN bond vectors, model (12) transforms to six-band Bloch Hamiltonian[79, 80] in momentum space with basis $\psi_k \equiv (a_{k,\uparrow}, b_{k,\uparrow}, c_{k,\uparrow}, a_{k,\downarrow}, b_{k,\downarrow}, c_{k,\downarrow})$

$$H(\mathbf{k}) = \begin{pmatrix} H_{3\times3}(\mathbf{k},\uparrow) & 0 \\ 0 & H_{3\times3}(\mathbf{k},\downarrow) \end{pmatrix} \quad (16)$$

where

$$H_{3\times3}(\mathbf{k},\eta_s) = \begin{pmatrix} \eta_s m_z & h_t^1(k)+\eta_s h_\lambda^1(k) & h_t^3(k)-\eta_s h_\lambda^3(k) \\ h_t^1(k)-\eta_s h_\lambda^1(k) & \eta_s m_z & h_t^2(k)+\eta_s h_\lambda^2(k) \\ h_t^3(k)+\eta_s h_\lambda^3(k) & h_t^2(k)-\eta_s h_\lambda^2(k) & \eta_s m_z \end{pmatrix} \quad (17)$$

Here $h_t^i(k) = -2t\cos(\mathbf{k}\cdot\mathbf{d}_i)$ and $h_\lambda^i(k) = 2i\lambda_0 \cos(\mathbf{k}\cdot\mathbf{v}_i)$. In the presence of inversion and time reversal symmetries, SOC opens a symmetric topological Dirac gap of $E_g = 4\sqrt{3}\lambda_0$. Like honeycomb lattice, Zeeman exchange interaction leads a topological phase transition from QSH phase to regular ferromagnetic metal such that two copies of spinful QAH phase are separated along energy axis. If Fermi level lie within Dirac gap of one of the spin sectors, it gives QAH effect with chiral edge state. Bulk band dispersion and corresponding energy spectrum for semi-infinite kagome nanoribbon are shown in Figure 5.

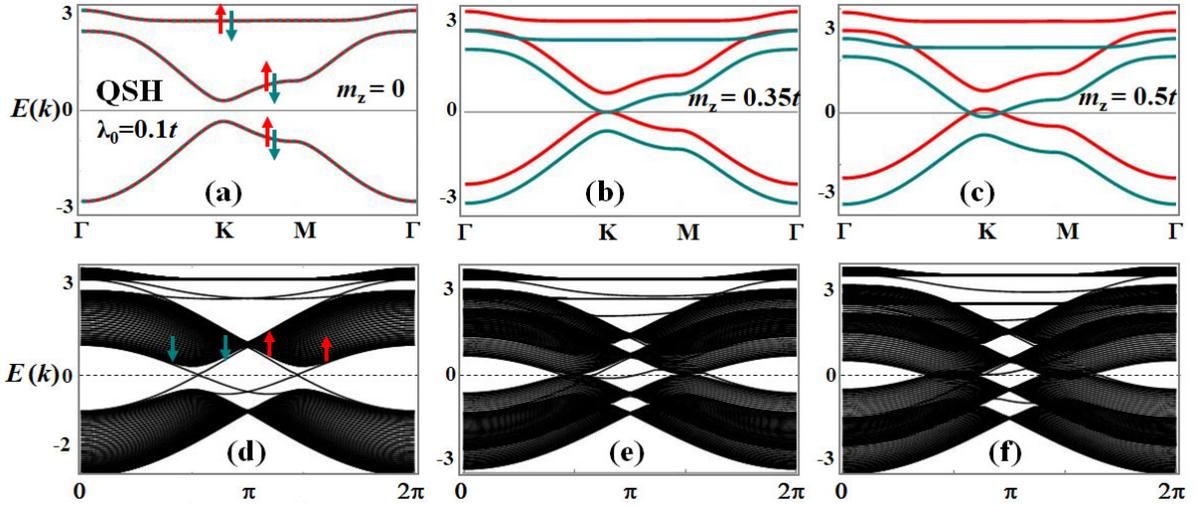

**Figure 5: Topological phase transition, driven by Zeeman exchange interaction, in QSH system with kagome lattice.** (*Top panel*) Bulk band structure for infinite sheet of kagome lattice showing topological phase transition driven by Zeeman exchange interaction. QSH phase **(a)** can be seen as two QAH-like spinful copies separated along energy axis where Zeeman exchange interaction turns the system into a regular ferromagnetic metal **(b, c)**. (*Bottom panel*) Corresponding energy dispersion for semi-infinite sheet of kagome lattice with zigzag edges showing helical edge states in QSH phase **(d)** and regular ferromagnetic metal **(e, f)**.

We conclude this section by noting that the topological phase transition driven by ferromagnetic exchange interaction in $(Bi,Sb)_2Te_3$ type thin films and QSH insulators with Kane-Mele type SOC is fundamentally different. Ferromagnetic $(Bi,Sb)_2Te_3$ type thin films host QAH effect whether the system is originally in topologically trivial or non-trivial insulating phase. On the other hand, any realistic Zeeman exchange interaction turns QSH insulator monolayers with Kane-Mele type SOC into a regular metal. However, as we discuss in next section, Kane-Mele type SOC can play an important role in materialization of QAH effect in ferromagnetic SGS with a signature of spinful Dirac dispersion in the absence of SOC.

## 6. QAH Effect in ferromagnetic spin-gapless semiconductors

Topological phase transition in Kane-Mele type spin-orbit-coupled QSH insulators, where spin is a good quantum number, persuades the possibility of QAH effect in ferromagnetic SGS. Hence, by setting electron spin polarization, say $s_z = -1$, Kane-Mele type SOC[77, 78] term effectively reduces to Haldane-like term for NNN tunnelling with complex hopping matrix elements[1]. Monolayers of transition-metal trihalides such as $MCl_3$ (M:V and Os)[41], $RuX_3$ (X: Br, Cl, I)[41, 42], $MnBr_3$[43], $NiCl_3$[44], transition-metal oxides $V_2O_3$[48], $Nb_2O_3$[49], $BaFe_2(PO_4)_2$[50] and transition-metal-organic framework $Mn_2C_{18}H_{12}$[51], $Co(C_{21}N_3H_{15})$[52], $TM(C_{18}H_{12}N_6)_2$[53] are typical ferromagnetic semiconducting materials where transition metal ions form honeycomb lattice structure. On the same footing, single layer $Cs_2Mn_3F_{12}$ of $Cs_2LiMn_3F_{12}$[55] and $Fe_3Sn_2$[54] are prototypical examples of ferromagnetic SGS where transition metal ions form kagome lattice structure.

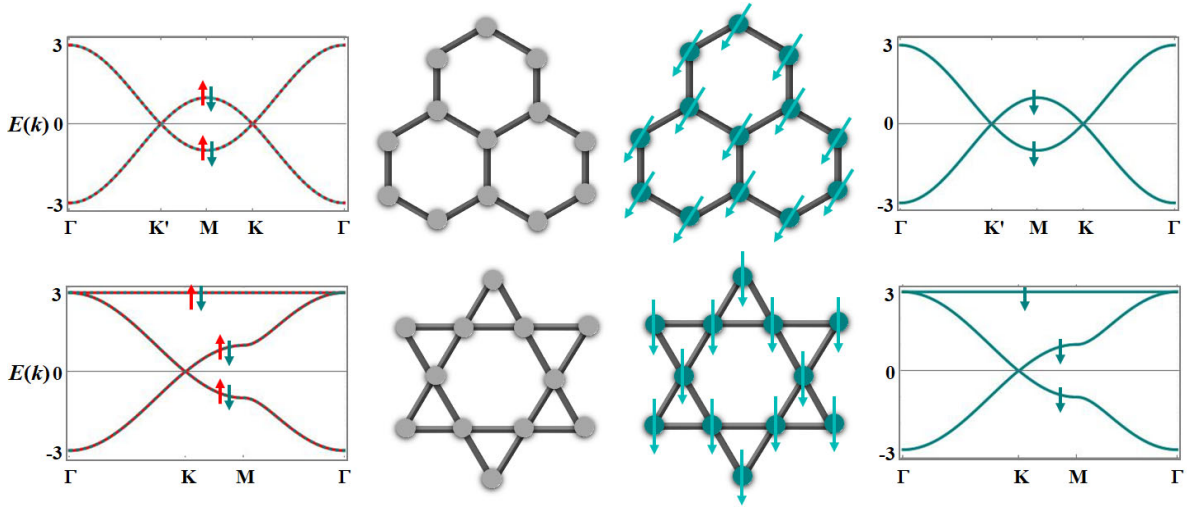

**Figure 6: Lattice structure for nonmagnetic (left) and ferromagnetic (right) honeycomb (top) and kagome (bottom) lattices.** In the absence of SOC, both honeycomb and kagome lattices with nonmagnetic (ferromagnetic) ground state host spin-degenerate (spinful) Dirac dispersion with four-fold (two-fold) Dirac points at the corners of Brillion zone. In the absence of SOC, ferromagnetic honeycomb/kagome lattices represent SGS where only one of the spin sectors represents spin-polarized Dirac dispersion while the other spin sector is gaped.

Ferromagnetic SGS can be modelled through two-band (three-band) spin-polarized tight binding Hamiltonian on a ferromagnetic honeycomb[51] (kagome[54, 55]) lattice by setting, say $s_z = -1$, in model (13) for QSH insulators

$$H = \varepsilon_0 \sum_i c_i^\dagger c_i - t \sum_{\langle ij \rangle} c_i^\dagger c_j - i\lambda_0 \sum_{\langle\langle ij \rangle\rangle} v_{ij} c_i^\dagger c_j \qquad (18)$$

In the absence of spinful NNN tunnelling (a single copy of Kane-Mele SOC), two-fold band crossings at K and K' represents spinful Dirac points in honeycomb (kagome) lattice. When NNN tunnelling is also activated, a spinful massive Dirac dispersion emerges with non-trivial topical gap. Band dispersion of infinite sheet and semi-infinite nanoribbon for honeycomb and kagome lattices confirms the non-zero Chern number as shown in Figure 7. In passing, the momentum space Berry curvature distribution and Chern number calculations for QAH model in ferromagnetic SGS with spinful Dirac dispersion are similar to that discussed for Haldane model.

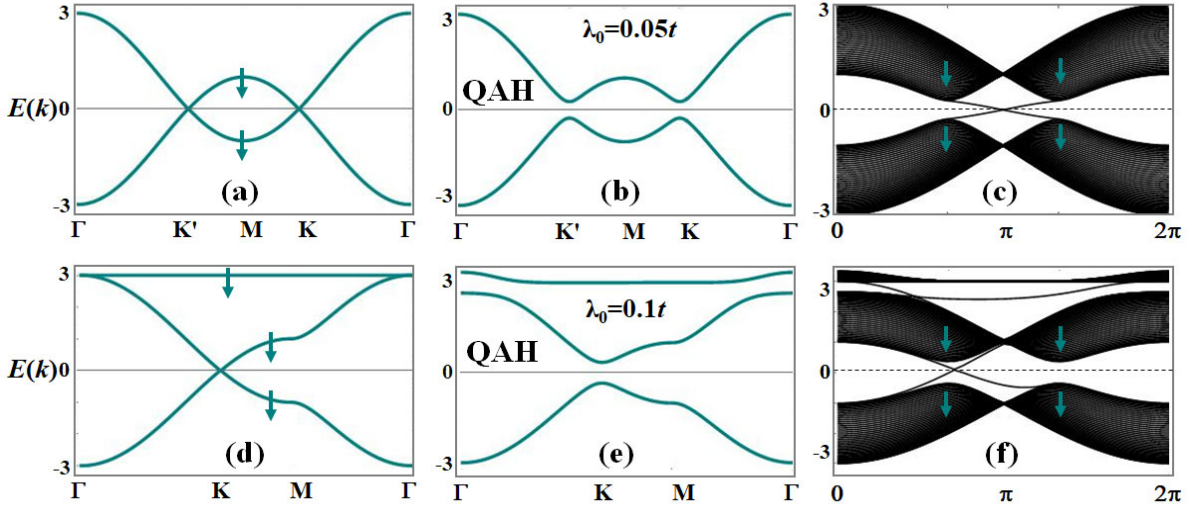

**Figure 7: Effects of spin-polarized Kane-Mele type SOC in SGS.** Bulk band structure and corresponding edge states for honeycomb lattice (*Top panel*) and kagome lattice (*Bottom panel*). With intrinsic ferromagnetic ground state, both honeycomb **(a)** and kagome **(d)** lattice host spin-polarized Dirac dispersion with two-fold Dirac points. When spinful Kane-Mele type SOC in two or three-band tight binding model (18) is considered, both ferromagnetic honeycomb and ferromagnetic kagome lattices leads to QAH effect **(b,e)**. Semi-infinite nanoribbon for honeycomb **(c)** and kagome **(f)** lattices with zigzag edges show chiral edge states.

## 7. Origin of Ferromagnetism in QAH Systems

Origin of exchange interaction in various QAH systems discussed above is mainly indebted to the partially filled *d*-manifolds of transition metal (TM) ions/adatoms/impurities. A large number of materials classes such as TM chalcogenides, TM halides, TM oxides, and various other families have been predicted to host intrinsic ferromagnetism[62]. In general, the nature of both magnetic and electronic properties depends upon various factors such as filling factors of *d*-manifold, nature/strength of bonding between TM cations (M) and anions (X) forming environment, nature of exchange interaction, and the lattice geometry in particular.

For example, in transition metal dihalides $MX_2$ monolayers where transition metal cation M = V, Cr, Mn, Fe, Co, Ni is surrounded by six halogen anions X = Br, Cl, I in octahedral environment, are simpler 2D magnetic systems. It has been repotred[89, 90] that $VX_2$, $CrX_2$, and $MnX_2$ monolayers are antiferromagnetic while $FeX_2$, $NiX_2$, $CoBr_2$ and $CoCl_2$ monolayers are ferromagnetic. Moreover, $FeX_2$ monolayers are conventional ferromagnetic half-metals[90, 91], $CoBr_2$ and $CoCl_2$ monolayers are ferromagnetic semiconductors with a very small band gap[92], while $NiX_2$ monolayers are ferromagnetic insulators[92]. Such diverse electronic structure is closely associated with spin configuration of *d*-manifolds: with high-spin configuration of $M^{+2}$ ($Fe^{+2}(3d^6)$, $Co^{+2}(3d^7)$, $Ni^{+2}(3d^8)$) ions in $MX_2$ monolayers, one of the spin sectors (say spin up) is distributed over whole *d*-manifold while the other spin sector (spin-down) fills the low-lying $t_{2g}$ orbitals with one, two, and three electrons in $FeX_2$, $CoX_2$, and $NiX_2$ monolayers respectively.

While localized magnetic moments arise through the competition between ionic and covalent bond strength between M and X ions generally, the magnetic ordering depends upon the interplay between two distinct types of exchange interaction: (i) the direct exchange between two NNN neighbours M-M ions, and (ii) the superexchange interaction between two NNN neighbours M-M ions mediated through NN X ions; M–X–M. Although both direct and superexchange leads to antiferromagnetic ordering in principle, Goodenough–Kanamori–Anderson (GKA) rules[93-95] rules indicate that the dominant superexchange interaction need not favour antiferromagnetic ordering in all cases: If *d*-orbitals on neighbouring M ions hybridize with orthogonal *p* orbitals on X ions such that M-X-M bond angles are $90^0$, the superexchange gives ferromagnetic ordering.

In addition, magnetic ordering can be controlled via external perturbations. For example, a detailed study of magnetic ordering in transition metal dichalcogenides (TMDC) monolayers and its tunability through strain can be found in reference[96]. In general, intra-layer M-X bonding is predominantly covalent in nature. However, since both ionic and covalent bonding is highly sensitive to the distance $d_{M-X}$ between M and X atoms, tensile (compressive) strain reduces (increases) the covalent bonding while increases (decreases) the ionic bonding interaction. In this context, large (small) $d_{M-X}$ results in large (small) unpaired spin accumulation on both M and X ions but reduces (enhances) the exchange interaction simultaneously. Impotently, M-M exchange interaction is affected relatively more than the M-X or M-X-M exchange interaction through variation of $d_{M-X}$ under tensile (compressive) strain.

## 7.1. Intrinsic ferromagnetism in 2D materials and the role of anisotropy

There are fundamentally important differences in both causes and the consequences of long-range magnetic order in the 3D and 2D structures. While magnetic order can easily be achieved in 3D structures, long-range magnetic order in 2D systems could be destroyed by thermal fluctuations, under Mermin–Wagner restriction[97]. However, 2D magnetic structures can reveal distinctive properties in contrast to their 3D counterparts. For example, Bonilla et al., showed that $VSe_2$ monolayer with 1T metallic phase on HOPG and $MoS_2$ substrate is ferromagnetic while 3D $VSe_2$ is paramagnetic[61]. It points out that, based on commonly employed technique of slicing down multilayered van der Waals (vdW) structures into monolayer structures, one cannot precisely establish the connection between the origin of magnetism in 2D monolayer crystal and their 3D bulk counterpart.

In order to precisely anticipate magnetic properties of 2D structures, fundamental understanding of the origin of magnetic ordering is desired rather than simply investigating their 3D bulk counterparts. Recent discovery of intrinsic ferromagnetism in two-dimensionally thin TM-compounds[59-62] indicates that the magnetic anisotropy is a highly desired feature of 2D magnetic structures. For example, the existence of ferromagnetism in $VSe_2$ monolayer with 1T metallic phase on HOPG and $MoS_2$ substrate is the consequence of strong in-plane anisotropy. This result is consistent with the predictions made through first principle simulations that both semiconducting 1H and metallic 1T phases of monolayer transition-metal dichalcogenides (TMDC) exhibit FM ordering with in-plane magnetization direction [96, 98, 99]. This strong magnetic anisotropy causing in-plane spin polarization in TMDC monolayers mainly originates from both through-bond and through-space interactions between transition metal and chalcogenide ions[96]. The experimental demonstration of intrinsic magnetism has already extended to various other transition metal compound monolayers such as itinerant ferromagnetism in 1T-$MnSe2$ [100, 101], $Fe_3GeTe_2$ [102], $Cr_2Ge_2Te_6$ [103], TM dihydride $MH_2$ (M=Sc,Co) [104] and electric field controlled magnetism in $Fe_3GeTe_2$ [105], $Cr_2Ge_2Te_6$ [103], and bi-layer $CrI_3$[106, 107].

## 7.2. Spin-gapless ferromagnets

Ferromagnetic spin gapless semiconductors are unique in the sense that the electronic properties are extremely sensitive to small perturbations – electrons can be easily transferred from occupied to empty states without overcoming any threshold energy barrier in the majority carrier gapless spin sector. As shown in figure 8, a comparison of density of states for metals with those of half-metal and spin-gapless semiconductors can be anticipated by electron spin polarization at Fermi level

$$P(E_F) = \frac{n_\uparrow(E_F) - n_\downarrow(E_F)}{n_\uparrow(E_F) + n_\downarrow(E_F)} \quad (19)$$

where $P = 0$ for paramagnetic metals with $n_\uparrow(E_F) = n_\downarrow(E_F)$, $0 < P < 1$ for ferromagnetic metals with $n_\uparrow(E_F) > n_\downarrow(E_F)$ while $P = 1$ for half-metals with $n_\uparrow(E_F) > 0$ and $n_\downarrow(E_F) = 0$. Similar to half-metals, minority spin channel is also gapped in spin-gapless material. However, the majority spin channel exhibits semi-metallic behaviour rather than metallic. As a result, electron spin polarization for spin-gapless materials is negligibly small which can be easily manipulated via external perturbations. While transport is mediated by both spin-up and spin-down electronic states in paramagnetic/ferromagnetic metals, only one kind of majority spin sates contribute in half-metallic/spin-gapless ferromagnetic transport phenomenon where minority spin channel remains gapped.

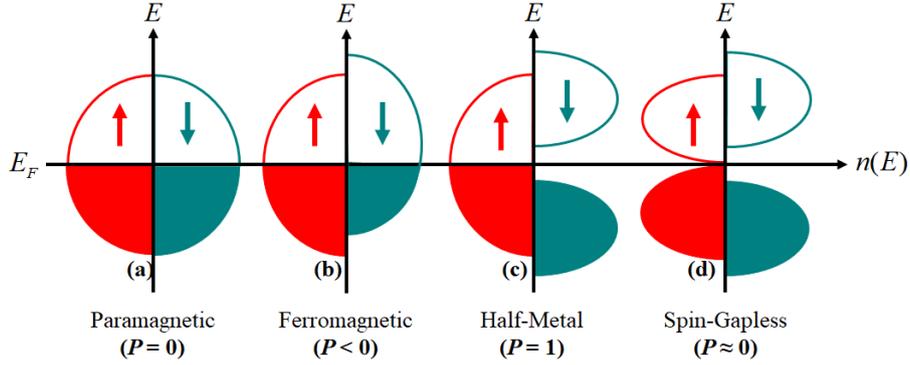

**Figure 8.** Schematics for density of states as a function of energy. Schematic representation of density of states for paramagnetic metals (a), ferromagnetic metals (b), ferromagnetic half-metals (c) and ferromagnetic spin-gapless semiconductors (d). Here red (cyan) arrows represent the majority (minority) spin states while filled (empty) regimes represent the occupied (empty) states.

## 7.3. Ferromagnetism in magnetic doped (Bi,Sb)$_2$Te$_3$ thin films

In diluted magnetic semiconductors thin films, where thickness of film is large enough such that conducting surface states exist explicitly, the ferromagnetism can be established through the Ruderman-Kittel-Kasuya-Yosida (RKKY) mechanism by doping magnetic impurities on the metallic surfaces or through Van Vleck paramagnetism by doping magnetic impurities in the insulating bulk. In diluted magnetic semiconductors doped with surface magnetic impurities[108], exchange coupling $H_{ex} = \sum_i J_i \mathbf{s}_i(\mathbf{r}) \cdot \mathbf{S}_i(\mathbf{r}) \delta(\mathbf{r} - \mathbf{R}_i)$ between itinerant electron spin density $\mathbf{s}_i(\mathbf{r}) = \psi^\dagger(\mathbf{r}) \sigma_i \psi(\mathbf{r})$ and spin of magnetic impurities $\mathbf{S}_i(\mathbf{r})$ at the location $\mathbf{r} = \mathbf{R}_i$ ($i=1,2,3,…..$), leads to the RKKY interaction among doped magnetic impurities $\mathbf{S}_1(\mathbf{r})$ and $\mathbf{S}_2(\mathbf{r}')$

$$H^{RKKY} = \sum_{i,j=x,y,z} J_{jk}(|\mathbf{r} - \mathbf{r}'|) \mathbf{S}_{1i}(\mathbf{r}) \mathbf{S}_{2j}(\mathbf{r}') \quad (20)$$

where $J_{jk}(|\mathbf{r}-\mathbf{r}'|)$ is the coupling constant between two magnetic impurities separated by $|\mathbf{r}-\mathbf{r}'|$. On the other hand, in dilute magnetic semiconductors doped with bulk magnetic impurities[2], a non-vanishing second order perturbation term in the ground state energy

$$\Delta E_n = \sum_{m \neq n} \frac{|\langle u_n(k)|\mu_B \mathbf{B}\cdot\mathbf{S}|u_m(k)\rangle|^2}{E_m - E_n} \quad (21)$$

leads to Van Vleck paramagnetic spin susceptibility for band insulators

$$\chi_z^{VV} = \frac{2\mu_0 \mu_B^2}{V} \sum_{m \neq n} \frac{|\langle u_n(k)|s_z|u_m(k)\rangle|^2}{E_m - E_n} \quad (22)$$

Here $u_n(k)$ and $u_m(k)$ are the Bloch wave functions for filled and empty bands with eigenenergies $E_m > E_n$, $\mu_0$ is the vacuum permeability, $\mu_B$ is Bohr magneton, and $s_z$ is the electron spin operator. The long range ferromagnetic order in 2D thin films of diluted tetradymite semiconductor family $Bi_2Se_3$, $Bi_2Te_3$, and $Sb_2Te_3$ has been experimentally realized – mainly due to the presence of both RKKY interaction and finite Van Vleck paramagnetic contribution – when doped with transition metal elements such as Fe[109], Cr[110], and V[111]. In summary, as experimentally demonstrated, Cr[9-12] and V[13] doped $(Bi,Sb)_2Te_3$ thin films provide ideal plate form for QAH effect.

In passing, it is intriguing to discuss the possible QAH phase in more conventional diluted magnetic semiconductors such as magnetically doped HgTe/CdTe type-I [3] and InAs/GaSb type-II [112] quantum wells. First of all, for the realization of Van Vleck paramagnetic electron susceptibility, the most relevant parameter is the hybridization gap generated through coupling between electron and hole sub-bands. As originally discussed via four-band models describing TI phase in both HgTe/CdTe type-I and InAs/GaSb type-II semiconducting quantum wells, hybridization gap in type-II QW is relatively small compared to that in type-I QW. This relatively small hybridization gap in magnetically doped type-II semiconducting quantum wells allows both RKKY interaction and Van Vleck paramagnetic electron susceptibility to contribute in ferromagnetic order [112]. Here, RKKY interaction is mediated through intraband charge flow due to small hybridization gap and band edge singularities while interband electron spin susceptibility originates Van Vleck paramagnetic.

On the other hand, experimental realization of QAHE in type-I semiconducting quantum wells does not seem plausible. While the itinerant electrons required to mediate the RKKY interaction between induced transition metal ions are missing, non-vanishing matrix elements of spin operator between the electron and hole bands is required to have a sizable susceptibility. However, unlike type-II semiconducting quantum wells, the ferromagnetic order via Van Vleck mechanism is hard to establish due to large hybridization gap in type-I semiconducting quantum wells. Secondly, unlike tetradymite semiconducting family $Bi_2Se_3$ where similar p-type electronic orbitals in both conduction and valance bands leads to a sizeable Van Vleck paramagnetic susceptibility, Van Vleck paramagnetic electron spin susceptibility is vanishingly small in type-I semiconducting quantum wells where conduction electron sub-bands are s-type and valance heavy hole sub-bands are p-type.

In summary, long range FM ordering can originate from both RKKY interaction as well as Van Vleck paramagnetic electron spin susceptibility in diluted magnetic semiconductors such as thin films of tetradymite semiconducting family $Bi_2Se_3$ and type-II semiconducting quantum wells InAs/GaSb but the contribution from both of these mechanisms in type-I semiconducting quantum wells vanishes. Finally, plausible QAHE in magnetically doped type-II semiconducting quantum wells is yet to be realized experimentally.

## 8. Discussion

Among several theoretical proposals made for the realization of QAH effect in various two-dimensionally thin materials, proposal for QAH effect in magnetic topological insulators are promising because of strong intrinsic SOC. QAH effect in impurity-induced topological insulator thin films[2] is the most successful model so far and it has already led to the experimental realization of QAH phase in Cr[9-12] and V[13] doped $(Bi,Sb)_2Te_3$ thin films. However, because of several experimental challenges, the highest temperature at which QAH phase is observed in these materials is extremely low. For example, non-uniform distribution of magnetic impurities can create spin scattering centres and hance, parallel dissipative channels at the one-dimensional edges are likely to induce several kΩ longitudinal resistance. Moreover, without affecting the band topology, high value of $T_c$ is difficult to achieve in diluted magnetic semiconductors $(Bi,Sb)_2Te_3$. As a result, 2D materials with quantized transverse resistance and extremely low longitudinal resistance are still awaiting.

Unlike $(Bi,Sb)_2Te_3$ thin films, QAH effect cannot be realized by inducing ferromagnetic exchange interaction in QSH insulators with Kane-Mele type SOC; any realistic Zeeman-like exchange interaction directly leads QSH insulator to regular ferromagnetic metal instead of QAH phase. However, despite this topological phase transition without entering QAH effect, Kane-Mele type SOC unveils another possibility: spinful Kane-Mele type NNN tunnelling can play an important role in materialization of quantum anomalous Hall effect in ferromagnetic SGS where the ground state is intrinsically ferromagnetic.

The possibility of QAH effect in ferromagnetic SGS with intrinsic ferromagnetism and SOC has an obvious advantage over impurity/proximity induced ferromagnetic QSH insulators – it can offer QAH phase without requiring random magnetic impurity doping or proximity effect. A suitable search of ferromagnetic semiconductors, where SOC is large and the ferromagnetic ground state is sustainable against thermal fluctuations at high temperature, can provide a platform for possible future dissipationless spintronics and low-energy electronics technologies.

**Acknowledgments**

This research is supported by the Australian Research Council (ARC) through ARC Professorial Future Fellowship project (FT130100778). We also acknowledge the support from Australian Research Council (ARC) Centre of Excellence in Future Low-Energy Electronics Technologies (FLEET Project No. CE170100039) and funded by the Australian Government.